# Time and Entropy in the Foundations of Mechanics

Roberto Assumpção
UNICAMP, Universidade Estadual de Campinas, SP, 13083-970, Brasil
UNIFAE, Faculdades Associadas de Ensino, S J B Vista, SP, 13870-377, Brasil
assump@fem.unicamp.br

*Abstract* - **This contribution analyses the classical laws of motion by means of an approach relating time and Entropy. Following the major conceptual issues introduced in the last century by the quantum and relativistic theories, that profoundly modified our understanding of mass, time and causality, we argue that adopting the notion of** *changes of states* **as opposed to the usual derivation of Newton´s laws in terms of** *fields***, a broader picture is obtained; for instance, description of the motion indicates that the act of experimentation disturbs the evolution of the system. Moreover, it is shown that a function of states can be written in terms of dynamical variables: masses and velocities. Association of this function to the concept of Entropy follows a procedure analogous to the derivation of the second law of Thermodynamics, giving rise to a unitary conceptual framework, suggesting an intriguing general picture where the diverse branches of mechanics turn out to be related by a common foundation.**

*Key Words*- Classical Mechanics, Entropy, Time

## INTRODUCTION

It has probably occurred to every student of the physical sciences that, apart from the inherent difficulties of deduction of Newton´s Laws [1–3], there is a lack of connection between the original formulation of Classical Mechanics and the subsequent development of the modern Quantum and Relativistic theories, particularly the former, which is formally based on the concept of "states of a system" rather than on the notion of times and fields. In fact, the formulation of the Classical Laws of Motion, by Sir Isaac Newton, marked the $17^{th}$ century transition from a philosophical qualitative description of observed facts to a physical quantitative representation of experimental results; in short, it was the birth of controlled laboratory methods. The primordial experiment of Mechanics [1] follows the procedure of measurement and repetition and is pictured in Figure 1: a body $m_1$ is abandoned near the proximity of a body M and the "rate of fall" determined by a time measurement; however, it is necessary to repeat the test, employing a distinct test body, say $m_2$.

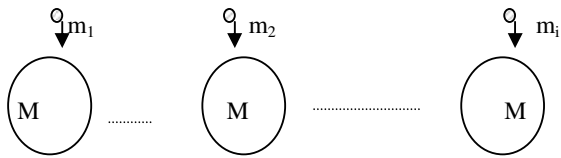

FIGURE 1 – The Classical Experiment

The result indicates a proportionality of the ratio of masses to the ratio of accelerations, that is:

$$\frac{m_1}{M} \propto \frac{\dot{V}}{\dot{v}_1} \quad \ldots\ldots \quad \frac{m_2}{M} \propto \frac{\dot{V}}{\dot{v}_2} \quad \ldots\ldots\ldots \quad \frac{m_i}{M} \propto \frac{\dot{V}}{\dot{v}_i}$$

where the pair ( $m_i, v_i$ ) refers to the "light" body whereas the pair ( M , V ) refers to the "heavy" body.
The general representation of the classical experiment is :

$$\frac{m_i}{m_j} = h \times \frac{\dot{v}_j}{\dot{v}_i} \quad (1)$$

where $h$, the "constant of proportionality", assumes the value $h = -1$ and, according to Classical mechanics, leads to the second ( $P_i = m_i v_i$ ) and to the third laws ( $F_i = -F_j$ ).

However, Figure 1 seems unrealistic once how could one generate, indefinitely, masses $m_i$ to experiment (balance) with M? A distinct view of the Classical experiment is shown in Figure 2.

Comparing Figures 1 and 2 we note that there is no place for $m_2$, other than on the surface of M, when $m_1$ is "falling". The same is true for any body $m_i$ used as a test body in the presence of M. Thus Figure 1 represents an imaginary or *gedanken* experiment, in a pure quantum-mechanical sense; that is, an experimental procedure that could not be carried out. In other words, the transition from the philosophical qualitative description of facts to a quantitative representation of experimental results was inconsistently conducted by Classical mechanics, leaving open *the problem of observation*.

Now Figure 2 pictures a 3-body problem which, as far as experimental methods are concerned, is the minimum necessary arrangement to assure consistent results; according to experimental data, the first and second stage gives, respectively:

$$\frac{m_1}{M+m_2} \approx \frac{\dot{V}_2}{\dot{v}_1} \quad ; \quad \frac{m_2}{M+m_1} \approx \frac{\dot{V}_1}{\dot{v}_2} \quad (2)$$

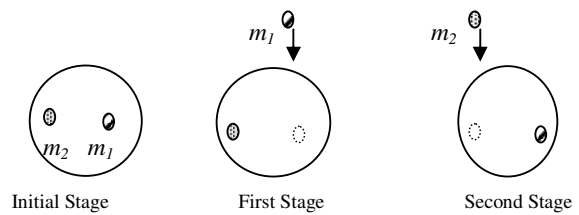

Initial Stage     First Stage     Second Stage

FIGURE 2 – Initial, first and second stages of a realistic Classical Experiment

Note that in the results of the experiment pictured in Figure 2, $\dot{V}_i$ is the acceleration of the conjunct $(M + m_i)$ and $\dot{v}_i$ the acceleration of $m_i$; the general representation of the experiment becomes:

$$\frac{m_1}{m_2} \approx \left\{ \frac{M + m_2}{M + m_1} \times \frac{\dot{V}_2}{\dot{V}_1} \right\} \times \frac{\dot{v}_2}{\dot{v}_1} \quad (3)$$

According to this relation, the 'constant of proportionality' $h$ between the ratio of masses and the ratio of accelerations assumes the value:

$$h \equiv \left\{ \frac{M + m_2}{M + m_1} \times \frac{\dot{V}_2}{\dot{V}_1} \right\} \quad (4)$$

Thus, the Classical "free fall" result becomes a 3–body experiment and $h$ reveals a dependency on the third part. A number of interpretations may result, but we note that a constant $h$, that is, independence on the masses $m_j$ indicates no repetition, or a result based on a single data; thus we will follow the possibility that the full value of $h$ can be obtained from the analysis of the experimental data.

## ANALYSIS OF THE "CLASSICAL" EXPERIMENT

Let $m_2 \approx m_1 + \delta M$ in (4); elimination of $m_2$ gives:

$$h \approx \frac{M + m_1 + \delta M}{M + m_1} \times \frac{\dot{V}_2}{\dot{V}_1} \quad (5)$$

or

$$h \approx \left(1 + \frac{\delta M}{M + m_1}\right) \times \frac{\dot{V}_2}{\dot{V}_1} \quad (6)$$

Now, taking $M \gg m_1$,

$$h \approx \left(1 + \frac{\delta M}{M}\right) \times \frac{\dot{V}_2}{\dot{V}_1} \quad (7)$$

Defining the ratio of masses $\alpha \equiv \frac{\delta M}{M}$, $h$ is given by:

$$h(\alpha) \approx (1 + \alpha) \times \frac{\dot{V}_2}{\dot{V}_1} \quad (8)$$

Conversely, elimination of $m_1$ instead of $m_2$ in equation (4) gives:

$$m_1 \approx m_2 - \delta M$$

so that $h$ becomes:

$$h \approx \frac{M + m_2}{M + m_2 - \delta M} \times \frac{\dot{V}_2}{\dot{V}_1} \quad (9)$$

that is, according to the previous definition of $\alpha$, we have:

$$h(\alpha) \approx \frac{1}{(1-\alpha)} \times \frac{\dot{V}_2}{\dot{V}_1} \quad (10)$$

Results represented by equations (8) and (10) appear contradictory; an analogous situation occurs when body 1 is considered heavier than body 2, that is:

$$m_1 \approx m_2 + \delta M$$

In fact, working out this last assumption, we are led with four distinct results for $h = h(\alpha)$. The additional values are:

$$h(\alpha) \approx \left(\frac{1}{1+\alpha}\right) \frac{\dot{V}_2}{\dot{V}_1} \quad (11)$$

$$h(\alpha) \approx (1 - \alpha) \frac{\dot{V}_2}{\dot{V}_1} \quad (12)$$

Relations (8), (10), (11) and (12) represent four distinct results for the "same" experiment; however, these are not precisely the same, once addition of a mass ($\delta M$) to $m_1$ is physically distinct from subtracting a mass ($\delta M$) from $m_2$.

Classical mechanics may argument that $h(\alpha)$ is constant, that all values converge to 1; moreover, a direct experiment relating bodies $(M + m_2)$ to $(M + m_1)$ is also virtual. However, the distinction of experiments is guaranteed by $\alpha$, not by employing different lesser bodies $m_1$ and $m_2$ in successive stages. A plot of these distinct results is given in Figure III.

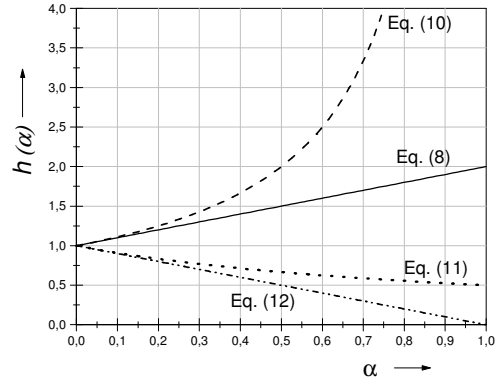

FIGURE 3 – Plot of the four distinct results of the Classical Experiment

The argument of the Quantum and Relativistic mechanics could be as follows: the first may argue that the particular value $h(\alpha)$ assumes is irrelevant, since this function is not an observable. Relativistic mechanics may admit such a function once a connection between masses and velocities do exist, but will argue that the effect can only be observed at relativistic velocities. Statistical mechanics, on the other hand, is the only theory that may consider the ambiguity and treat $h(\alpha)$ as a description of the path followed by the masses, a function that is sensitive to the order of experimentation and takes the responsibility for the fact that ideal isolation of the two bodies $m_1$ and $m_2$ could not be achieved. This last argument will now be followed, in a study of the evolution of the mechanical states of the system [M, $m_1$, $m_2$].

## EVOLUTION OF THE STATES OF THE SYSTEM

Historically, the subject of thermodynamics arose before the atomic structure of matter was understood [4,5]. Macroscopic systems began to be studied from a phenomenological point of view in the 19$^{th}$ century, giving rise to the laws of thermodynamics and the concept of Entropy as a function that characterizes each *macrostate* of a system. When the theory of atomic constitution of matter gained importance, macroscopic systems began to be analyzed from a microscopic point of view as systems consisting of very many particles or subsystems.

This *atomic approach* to macroscopic problems, originally from the kinetic theory of gases, was then consistently developed (among many other contributors) by Clausius – the man who *invented* Entropy, Maxwell – that formulated the distribution law of molecular velocities and Boltzmann, with his fundamental equation in an integrodifferential form. The work of Boltzmann [6] provided a more general and fundamental analyses of *irreversible process* and *the approach to equilibrium*, giving rise to the further development of the discipline of *statistical mechanics*, that is, the method of calculus of the macroscopic parameters of a system from a knowledge of its microscopic constituents.

In the sequence of this contribution our attempt is to follow this method, as if the *tripod* – Clausius, Maxwell and Boltzmann arrived before Newton.

Figure 4 pictures the evolution of the masses M, $m_1$ and $m_2$. In opposition to the usual formulation of the "free fall" analysis, at least two times can be distinguished: the relaxation time of the bodies $m_1$ and $m_2$ when taken outside M, or the elapsed time of "fall" ($\delta t_i$) and the time interval between observations, or the rate of change of the whole mechanical system ($\delta t$). There is also a third time associated to the body M – $\Delta t$, named the *proper time* of the body. $\Delta t$ is also the proper time of $m_1$ and $m_2$ when these are simultaneous with M, such as in stage $S_o$.

Defining the state S of the system by the masses present at a particular time, the initial state can be represented by:

$$S_o \equiv \frac{M + m_1 + m_2}{\Delta t} \qquad (13)$$

Following this definition, the stages of the mechanical system [ M, $m_1$ , $m_2$ ] can be represented by similar relations:

$$S_1 \equiv \frac{M + m_2}{\Delta t + \delta t} + \frac{m_1}{-\delta t_1} \quad S_2 \equiv \frac{M + m_1 + m_2}{\Delta t + 2\delta t}$$

$$S_3 \equiv \frac{M + m_1}{\Delta t + 3\delta t} + \frac{m_2}{-\delta t_2} \quad S_4 \equiv \frac{M + m_1 + m_2}{\Delta t + 4\delta t} \qquad (14)$$

At the initial stage $S_o$ all the bodies are together, but at the first experimentation stage, $S_1$, the body $m_1$ is retired from contact with M and $m_2$. Then $m_1$ follows its own way and lasts a time – $\delta t_1$ to reach equilibrium. An analogous situation appears in stage 3 for the body $m_2$. The negative value indicates that these bodies are relaxing back to $\Delta t$, or to equilibrium. Due to the evolution of the system, the falling bodies reach equilibrium at a later time, $\Delta t + n\delta t$ (n = 1,2,...), where ($\Delta t + n\delta t$) is the proper time of the simultaneous bodies and $\delta t$ the time interval between observations. Note that $\delta t \geq \delta t_i$ is an experimental condition imposed by the nature of the measurement process: the evolution of the mechanical system must be lower than the "rate of fall" of the (internal) bodies $m_1$ and $m_2$.

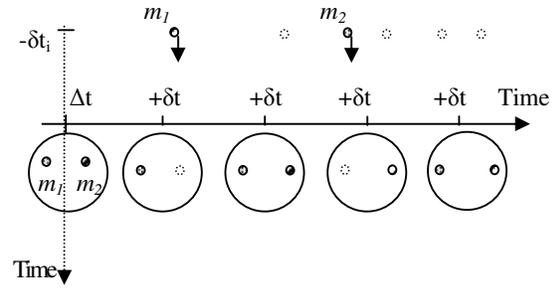

FIGURE 4 – Time Evolution of a three body Mechanical System

This last condition is a constraint, representing an uncertainty on the observation of $\delta t$ carried out by the (internal) observer that measures $\delta t_1$ and $\delta t_2$. Here we note that from the point of view of Newtonian Classical Mechanics this constraint is irrelevant, once observer measures only the time of fall $\delta t_i$ of the bodies $m_i$. The body M, say, the Earth, is not taken into account.

Considering Equations (13) and (14), the changes between adjacent steps can be written as :

$$\Delta S_{i,i+1} \equiv S_{i+1} - S_i \qquad (15)$$

Conversely, Figure 4 shows a dynamical evolution of the system [M, $m_1$, $m_2$] whereas each stage represents a static picture of M, $m_1$ and $m_2$; thus, as a result of evolution, each stage can also be represented as:

$$S_i^* \equiv S_i + \Delta S_{i-1,i} \qquad (16)$$

$S_i^*$ can be computed for all stages (i >0) and represents an excited or meta-stable state resulting from the act of experimentation, as opposed to the static picture. Now, as a result of the distinct values of $h(\alpha)$ ( Figure 3) two ways of evolution can be distinguished: evolution "A", when the first experiment is the "free fall" of body $m_1$ and the second the "free fall" of body $m_2$ and evolution "B" when the first experiment is the "free fall" of body $m_2$ and the second the "free fall" of body $m_1$. Note that distinction in this case relates to the order of experimentation, leading to a reading of the states S as a mechanical Entropy of the whole system [M, $m_1$, $m_2$]. We compute the changes for one of these two cases and present the result for the other evolution.

Following the picture of Figure 4 and computing the changes, we have for the first evolution from stages 0 → 1:

$$\Delta S_{01} \cong \frac{M+m_2}{\Delta t + \delta t} - \frac{m_1}{\delta t_1} - \frac{M+m_1+m_2}{\Delta t} \quad (17)$$

rearranging the terms,

$$\Delta S_{01} \cong (M+m_2)\left[\frac{1}{\Delta t + \delta t} - \frac{1}{\Delta t}\right] - m_1\left(\frac{1}{\Delta t} + \frac{1}{\delta t_1}\right)$$

Furnishing the first stage $S_1^*$ after experimentation:

$$S_1^* \cong (M+m_2)\left[\frac{2}{\Delta t + \delta t} - \frac{1}{\Delta t}\right] - m_1\left(\frac{1}{\Delta t} + \frac{2}{\delta t_1}\right) \quad (18)$$

Similarly, stages 2 and 3 after experimentation are:

$$S_2^* \cong (M+m_2)\left[\frac{1}{\Delta t} + \frac{2}{\Delta t + 2\delta t} - \frac{2}{\Delta t + \delta t}\right] + m_1\left(\frac{1}{\Delta t} + \frac{2}{\Delta t + 2\delta t} + \frac{2}{\delta t_1}\right) \quad 19)$$

$$S_3^* \cong -M\left(\frac{1}{\Delta t} - \frac{2}{\Delta t + \delta t} + \frac{2}{\Delta t + 2\delta t} - \frac{2}{\Delta t + 3\delta t}\right) -$$

$$- m_1\left(\frac{1}{\Delta t} + \frac{2}{\Delta t + 2\delta t} - \frac{2}{\Delta t + 3\delta t} + \frac{2}{\delta t_1}\right)$$

$$- m_2\left(\frac{1}{\Delta t} - \frac{2}{\Delta t + \delta t} + \frac{2}{\Delta t + 2\delta t} - \frac{2}{\delta t_2}\right) \quad (20)$$

Thus we call final stage of the evolution pictured in Figure 4 as $S^A$, in order to explicitly consider the difference in the order of evolution of the system; thus, $S^A = S^*_4$, is:

$$S^A \cong (M+m_1+m_2)\left\{\frac{1}{\Delta t} + \frac{2}{\Delta t + 2\delta t} + \frac{2}{\Delta t + 4\delta t}\right\} + 2\frac{m_1}{\delta t_1} + 2\frac{m_2}{\delta t_2} -$$

$$- (M+m_2)\frac{2}{\Delta t + \delta t} - (M+m_1)\frac{2}{\Delta t + 3\delta t} \quad (21)$$

Alternatively, the order of experiments can be reverted by exchanging $m_1$ and $m_2$; states $S_o$, $S_2$ and $S_4$ are unchanged, but stages 1 and 3 become:

$$S_1 \equiv \frac{M+m_1}{\Delta t + \delta t} + \frac{m_2}{-\delta t_2} \;;\; S_3 \equiv \frac{M+m_2}{\Delta t + 3\delta t} + \frac{m_1}{-\delta t_1} \quad (22)$$

This leads to a final state slightly different from $S^A$ given by Eq. (21); calling this other evolution as $S^B$, we have:

$$S^B \cong (M+m_1+m_2)\left\{\frac{1}{\Delta t} + \frac{2}{\Delta t + 2\delta t} + \frac{2}{\Delta t + 4\delta t}\right\} + 2\frac{m_1}{\delta t_1} + 2\frac{m_2}{\delta t_2} -$$

$$- (M+m_1)\frac{2}{\Delta t + \delta t} - (M+m_2)\frac{2}{\Delta t + 3\delta t} \quad (23)$$

This shows that, depending on the order of experimentation, the final state of the system differs. Defining a quantity $\Delta S_P = |S^A - S^B|$ representing the variation associated to the process, we have:

$$\Delta S_P \cong 2(m_1 - m_2)\left[\frac{1}{\Delta t + \delta t} - \frac{1}{\Delta t + 3\delta t}\right] \quad (24)$$

This result is independent of the choice of the light bodies, since no matter $m_1 = m_2 + \delta M$ or $m_2 = m_1 + \delta M$, the value of $\Delta S$ remains the same. Thus experimentation gives rise to distinct results, according to the order of the procedure. Conversely, the act of experimentation disturbs the system, in a sense that it does not return to the same configuration; the changes related to each of the bodies M, $m_1$ and $m_2$ can also be calculated. Choosing the last evolution, the accumulated change for each one of the bodies can be computed.

**RESULTS AND DISCUSSION**

The change related to body M is:

$$\Delta S_M \cong M\left(\frac{1}{\Delta t} - \frac{2}{\Delta t + \delta t} + \frac{2}{\Delta t + 2\delta t} - \frac{2}{\Delta t + 3\delta t} + \frac{1}{\Delta t + 4\delta t}\right)$$

This result can be written as two sets,

$$\Delta S_M \cong M\left(\frac{1}{\Delta t} - \frac{1}{\Delta t + \delta t} + \frac{1}{\Delta t + 2\delta t} - \frac{1}{\Delta t + 3\delta t} + \frac{1}{\Delta t + 4\delta t}\right) -$$

$$- M\left(\frac{1}{\Delta t + \delta t} - \frac{1}{\Delta t + 2\delta t} + \frac{1}{\Delta t + 3\delta t}\right)$$

According to Leibniz's theorem [7] for alternating sets, each one gives a positive sum not superior to the first term, that is:

$$0 \leq \Delta S_M \leq M\left(\frac{1}{\Delta t} - \frac{1}{\Delta t + \delta t}\right)$$

Thus a rough result is:

$$\Delta S_M \cong M\left(\frac{1}{\Delta t} - \frac{1}{\Delta t + \delta t}\right) \quad (25)$$

Similarly, the resulting change for the bodies *m₁* and *m₂* are:

$$0 \leq \Delta S_{m1} \leq \frac{2m_1}{\delta t_1} \quad (26)$$

$$0 \leq \Delta S_{m2} \leq \frac{2m_2}{\delta t_2} \quad (27)$$

Taking the right hand of Equations (26) and (27) we can write:

$$\frac{\delta t_1}{2} \Delta S_{m1} \leq m_1 \quad (26')$$

$$\frac{\delta t_2}{2} \Delta S_{m2} \leq m_2 \quad (27')$$

Recovering that the quotient $m_1/m_2$ was already determined by the Classical experiment, Figure 2, furnishing the usual ratio of accelerations – Equations (1), (3), the last two relations give:

$$\frac{m_1}{m_2} \cong h(\alpha) \frac{\dot{v}_2}{\dot{v}_1} \geq \frac{\Delta S_{m2} \delta t_1}{\Delta S_{m1} \delta t_2} \quad (28)$$

Here we note that $h(\alpha)$, defined by Equation (4) as a function of masses of the three bodies (M, $m_1$, $m_2$) and the acceleration of body M ($\dot{V}_i$) plays the same role as the Newtonian constant of proportionality between masses and accelerations and, apart from the existence of distinct values, as given by Equations (8), (10), (11) and (12), it is still essential for the classical definition of masses.

On the opposite, elimination of the times in Eq. (28) by means of the usual definition of acceleration, that is:

$$\langle v_i \rangle \equiv \delta t_i \dot{v}_i \quad (29)$$

and taking $<v_i>$ as the medium value of the velocity acquired by body $m_i$, we can finally establish a relation among masses (embodied in $h(\alpha)$), velocities, and the change of states of the Mechanical System throughout the experimental procedure, as represented by ΔS :

$$\frac{\Delta S_{m_1}}{\Delta S_{m_2}} \approx h(\alpha) \frac{\langle v_2 \rangle}{\langle v_1 \rangle} \quad (30)$$

Finally we note that Equation (30) has a form similar to equation (1): velocities take the place of accelerations and the ratio of masses are substituted by the "ratio of states"; also, the motion of the [$m_1$ | $m_2$] subsystem probably occurs in the direction of minimum relative ΔS changes. Moreover, though equations (26') and (27') were written in terms of the variables of the bodies $m_1$ and $m_2$, equation (30) inserts the *h* function as a connection with the motion of the system as a whole; this opens the possibility to interpret the different values of *h* as distinct paths that the light bodies ($m_1$, $m_2$) may follow on their way back to equilibrium, a "classical" uncertainty in the motion of the mechanical system. Conversely, this function is directly related to the picture of Newtonian mechanics represented by Figure 2, so the explicit introduction of the time of the system Δt in the Classical formalism allowed the description of the motion in terms of change of states without any significant modification of the Newtonian results.

## CONCLUSIONS

This contribution reviews the original experiment of Classical Mechanics centering the discussion on the evolution of states and explicitly pointing the idea that the experimental procedure (Fig. 1) that leads to the definition of masses and derivation of the "laws of motion" is unrealistic. Analysis of a realistic experiment (Fig.2) leads to multiple solutions, expressed by the values of $h(\alpha)$, revealing that the classical results are based on a particular result ($\alpha \rightarrow 0$). Application of a simple description of the motion, in terms of change of stages, without any attempt to quantification other than the original Newtonian results, indicates that the act of experimentation disturbs the evolution of the system.

Employing the concept of Mechanical States, those distinct results can be associated to variations of the order of the whole mechanical system [ M, $m_1$, $m_2$]. This opens the possibility to the introduction of a function "S", defined in terms of the dynamical variables of the system, that takes into account the process of change "ΔS" of the mechanical system, i.e., the condition under which the process occur and when the system is in equilibrium.

According to the results obtained, this function applies to any realistic process taking place in an isolated system, such as the one described here – the "free fall", though a number of other classical systems, such as the motions of a simple pendulum, the harmonic oscillations of a body connected to a spring, etc., could be provided following the same perspective.

This leads to a plausible association of this function to the concept of Entropy, as first devised by Clausius, in the context of Thermodynamics and the subsequent development of the balance between macroscopic and microscopic methods of approach to equilibrium of mechanical systems.

This suggests that the original Newtonian picture of Classical Mechanics expressed in terms of times (and fields) can be adapted to a broader view more compatible with the further development of mechanics, as put forward throughout the last century scientific revolution, embodied in the relativistic and quantum theories, giving rise to a unitary and comprehensive conceptual framework for the mechanical sciences. The current approach seems compatible with the contemporary efforts [8–10] on recovering and bringing the fundamentals of the basic sciences in to Engineering curricula.

## ACKNOWLEDGMENT

Useful discussions with W Amstalden are gratefully acknowledged.